# PV CEP AND V350 CEP: STARS ON THE WAY BETWEEN FUORS AND EXORS

H. R. Andreasyan,[1] T. Yu. Magakian,[1] T. A. Movsessian,[1] and A. V. Moiseev[2]

*Based on new observations during 2015-2020 and published data, the unusual eruptive variables PV Cep and V350 Cep are examined. It is shown that PV Cep underwent a regular outburst followed by a drop in brightness that lasted overall from 2011 to 2019 and is still in a deep minimum. The outburst was accompanied by substantial changes in the intensity and profiles of a number of lines, including Hα, [SII], and [OI]. The forbidden lines generally have negative radial velocities and can be divided into four components, with variable velocities and relative intensities. V350 Cep essentially is at a maximum brightness level over the entire time and its spectrum is practically unaltered. The available data suggest that the pronounced P Cyg profile of the Hα line in the spectrum of V350 Cep appeared several years after the luminosity rise, in 1986. The luminosities of the stars in the current state are estimated to be 20 $L_\odot$ and 3.3 $L_\odot$, respectively. It is concluded that both stars may srepresent a so-called intermediate objects between the FUor and EXor classes.*

## 1. Introduction

The thousands of young low-mass ($M < 2 M_\odot$) stars referred to as T Tau stars that have been discovered up to now are divided into two main subclasses:

[1]V. A. Ambartsumian Byurakan Astrophysical Observatory, National Academy of Sciences of Armenia; e-mail: hasmik.andreasyan@gmail.com
[2]Special Astrophysical Observatory (SAO), Russian Academy of Sciences

– classical T Tau stars, or CTTS (with strong optical variability, a developed emission spectrum, and, often, distinct signs of a collimated outflow) and

– "weak" T Tau stars, or WTTS (low brightness variations, an emission spectrum often limited to the emission of H$\alpha$, matter outflow rarely observed).

The difference between these subclasses also shows up in other regions of the spectrum. According to the most widespread viewpoint, mostly confirmed observationally, the factor determining to which of the two classes a star belongs is the rate of magnetohydrodynamic accretion of matter from a circumstellar disk onto the star.

Eruptive young stars, which exhibit sharp and sometimes extremely prolonged rises in brightness accompanied by spectral changes and the ejection of matter, are of greatest interest for understanding this phenomenon. Of these the best known and rarest (less than twenty are known at present) are FU Ori objects (FUors), for which the time of maximum brightness may last thousands of years according to statistical estimates. Other, also extremely rare and interesting classes are the EX Lup objects (EXors) and the so-called intermediate objects. Although their rises in brightness (not as prolonged as in FUors) are usually explained by changes in the accretion rate, the details and mechanisms of this phenomenon are not yet clear, in part because of the small number of observed examples. Thus, it is extremely important to extend and supplement the existing information on eruptive stars in their early stage of evolution.

This paper is devoted to further study of two known eruptive stars PV Cep and V350 Cep, for which the nature of their variability has not yet been fully clarified.

The variability of the star PV Cep and the associated cometary nebula GM1-29 (RNO 125) was discovered in 1997 [1,2]. The sudden rise in the luminosity of PV Cep resembled a FUor outburst, but the very first spectra of this star appeared similar to classical T Tau stars [2-4]. Herbig subsequently assigned PV Cep to the EXor class [5] which he had introduced, but he later eliminated this object from the list [6] since further studies revealed too many significant differences between PV Cep and typical EXors. We note that in the subsequent years PV Cep has undergone several more rises and falls in brightness (see below). Like many other active T Tau stars, it is the source of an extended Herbig-Haro flow [7,8] and of a bipolar molecular outflow [9]. Despite a small number of subsequent studies, of which Refs. 10-13 should be noted and will be examined below, in many regards PV Cep is still an enigmatic object.

The variable star V350 Cep which is located in a cluster inside the nebula NGC 7129, was first mentioned in Ref. 14. It was pointed out that this object was not visible in the charts of the Palomar Atlas (1954), and had reached a brightness of $16^m.5$ (V) by 1977. A historical light curve of V350 Cep was presented in Ref. 15. Later data were collected by Herbig [6]. These results show uniquely that the rise in brightness of this star began roughly in 1971-72; in 1978, having reached a maximum, the star has remained at the same level for almost 50 years, with only weak fluctuations. Twice (in 2009 and 2016) deeper minima (to $\Delta V = 1^m.77$) were observed lasting several months followed by a return of the brightness to its average value [16,17].

However, even V350 Cep turned out not to be a FUor, at least a classical one, since the first spectral observations showed that it is a T Tau star with a strong and highly developed emission spectrum [3], as confirmed subsequently by numerous observations, e.g., Refs. 18-20. All authors have noted the emission character of the spectrum of V350 Cep with strong Balmer lines. [OI] forbidden lines could also be seen, along with many lines of both ionized and neutral iron. A detailed analysis of V350 Cep with high resolution spectrographs [6,21] revealed, in particular, obvious signs of the outflow of matter from the star. Herbig [6] pointed out that V350 Cep could hardly be assigned even to the EXors, since, in particular, this star did not manifest the characteristic (for the latter) outbursts and is essentially constantly at the maximum brightness level. The spectral class of V350 Cep was ultimately fairly reliably assessed as M2 [18,6].

In this paper we take new spectra of PV Cep and V350 Cep obtained over the last five years and compare these results with other data.

## 2. Observations

Observations of both objects were made on the 2.6-m telescope at the Byurakan Astrophysical Observatory. We used a SCORPIO spectral camera at the principal focus of the telescope to obtain both direct images and long-slit spectra. An analogous camera is also installed at the 6-m BTA telescope at the Special Astrophysical Observatory (SAO) of the Russian Academy of Sciences [22]. Initially a TK SI-003A 1044×1044 CCD matrix was used as a detector and after August 2016, an E2V CCD42-40 2080×2080 matrix. When taking spectra, the slit width was 1".5 with a length of about 5'. The field of view for taking direct images was about 11' for a scale of 0.67"/pixel. The dispersing elements were volume-phase holographic lattices with 600 and 1800 g/mm, providing spectral resolutions R of roughly 800 and 2500, respectively.

For calibration an Ne+Ar lamp was used for comparison spectra. The overall exposure time was calculated so that a signal/noise ratio greater than 100 resulted in the spectra following processing and optimal extraction.

Table 1 is a summary of all the spectral observations of PV Cep at Byurakan and the SAO. At the 6-m telescope at the SAO, a SCORPIO-2 reducer at the primary focus [23] and an E2V 261-84 CCD as the one spectrum was obtained yielded a single on Dec. 23, 2020. The width of the spectrograph slit was 1" with a length of about 6'. A single spectrum of V350 Cep was obtained at Byurakan with a total exposure of 20 min and a resolution of about 2500.

The ESO-MIDAS program was used to process the spectra from the 2.6-m telescope and the SCORPIO-2 data were processed in the IDL environment using programs developed at the SAO. For both stars the equivalent widths and radial velocities of all the main lines were measured. For convenience in comparison the spectra of PV Cep were normalized to the continuum.

TABLE 1. Observation Log for PV Cep on the 2.6-m and 6-m Telescopes

| Observation date (d.m.y.) | Spectral range (Å) | Resolution ($\lambda/\Delta\lambda$) | Total exposure (min) |
|---|---|---|---|
| 20.09.2015 | 4100-6800 | 800 | 30 |
| 10.06.2016 | 5800-6800 | 2500 | 40 |
| 24.08.2016 | 5800-6800 | 2500 | 15 |
| 21.12.2016 | 5800-6900 | 2500 | 45 |
| 24.06.2017 | 5800-6900 | 2500 | 40 |
| 25.06.2017 | 5800-6900 | 2500 | 40 |
| 20.06.2018 | 5800-6800 | 2500 | 60 |
| 23.12.2020 | 3650-7300 | 1300 | 40 |

**3. Results and discussion**

**3.1. Photometric data and the light curve of PV Cep.** Although we periodically obtained direct images of PV Cep on which strong variations in the star's brightness and changes in the shape of the associated nebula GM 1-29 (RNO 125) showed up clearly, no photometric estimates of the brightness were made, since this star is under constant monitoring by the AAVSO group. The AAVSO data concerning the PV Cep starts from 2010, and photometry for 2004-2010 is described in Ref. 12.

An examination of the summary light curve for PV Cep plotted on the basis of these data shows that since 2006 the star began to fall in brightness from $R \approx 13^m.5$ to $R \approx 15^m.5$, and after 2008 fell rapidly to a minimum ($R \approx 16^m.5 - 17^m.4$). During 2010, the brightness of PV Cep increased in waves and then fell in amplitude by roughly $1^m$-$1^m.5$. Finally, essentially since the beginning of 2011, the star began gradually and almost continuously to become brighter, practically reaching a maximum ($R \approx 13^m.6$) toward the end of 2016. After this, a rapid drop in brightness began with superimposed individual local, brief minima (an especially noticeable example occurred in autumn 2017). Toward the middle of 2019 the star finally went into a minimum ($R \approx 17^m - 17^m.3$), where it remains up to the present (the end of 2020). Thus, over the last 15 years PV Cep has undergone two large-scale (in amplitude, $3^m$-$4^m$, and in duration, several years) rises in brightness. An analysis of older published data reveals other maxima over 1950-2005 (Kun, et al., unpublished), but it may be suspected that PV Cep reached the maximum brightness (of order $R = 11^m$) only during the most powerful eruption of 1976-78, when it was discovered [4].

Figure 1 shows an R-band light curve of PV Cep for 2015-2020, generated from AAVSO data, where the arrows indicate the dates of our observations. The AAVSO curve breaks off before the end of 2020, when the final spectrum was obtained, but our observations confirm that PV Cep stayed in a deep minimum (R=$17^m.3$-$17^m.5$) throughout 2020.

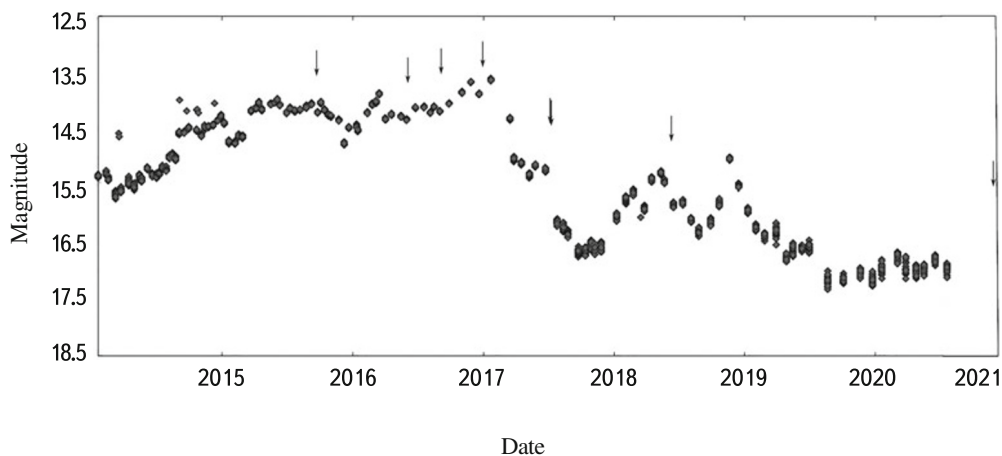

Fig. 1. An R-band light curve of PV Cep. The arrows indicate days on which spectra were taken.

**3.2. General character of the spectrum of PV Cep and its variations.** In general, the spectrum of PV Cep in the visible and near infrared has been described in detail [4,10,12,13]. These results can be summarized briefly as follows. The star has a spectrum that is typical for an extremely active T Tau stars: very strong emission in the Hα line, as well as noticeable emission in the Paschen and Brackett series, the CaII IR triplet, numerous FeII and FeI lines. In IR range can be seen lines of many other ions. NaD absorption lines are very prominent. Lines with the highest excitation level are not strong, but distinctly noticeable λ5875 and λ6678 HeI. In addition, a small number of forbidden lines of shock excitation are observed in the spectrum of PV Cep: in the optical range, [OI] and [SII] emission, and in the near IR, numerous [FeII].

It should be noted that the star has a very red color, so that the short-wavelength part of the visible range was rarely observed. Thus, the strong emission K line of CaII and the fluorescence in the FeI lines typical of the most active T Tau stars [10] observed during the epoch of the highest rise in brightness (1978) remain the only ones of their kind. We shall examine the question of the spectral class of PV Cep separately. Earlier studies have also revealed substantial variability of individual spectral lines apparently related to the brightness level of the star.

Comparing Table 1 and the light curve of PV Cep, we see that our spectra correspond to essentially all levels of brightness of the star. For comparison, in Fig. 2 we show the spectrum of PV Cep at a brightness maximum and minimum. No substantial changes occurred compared to the data of other authors; at the same time, the differences between the two spectra are perfectly obvious. The first thing that comes before the eyes is the much greater intensity of the forbidden lines, while the classical set of FeI and FeII emission undergoes no particular changes. In particular, despite the weakened continuum level, it can be seen clearly that in the blue-green ($\lambda\lambda 5000$-$5500$ Å) region of the spectrum the structure of the "forest" of numerous metal lines is essentially the same. We note that the intensity of the $\lambda 5875$ HeI line also remains almost unchanged; i.e., the excitation level in the chromosphere of the star is also persisted during the brightness oscillations.

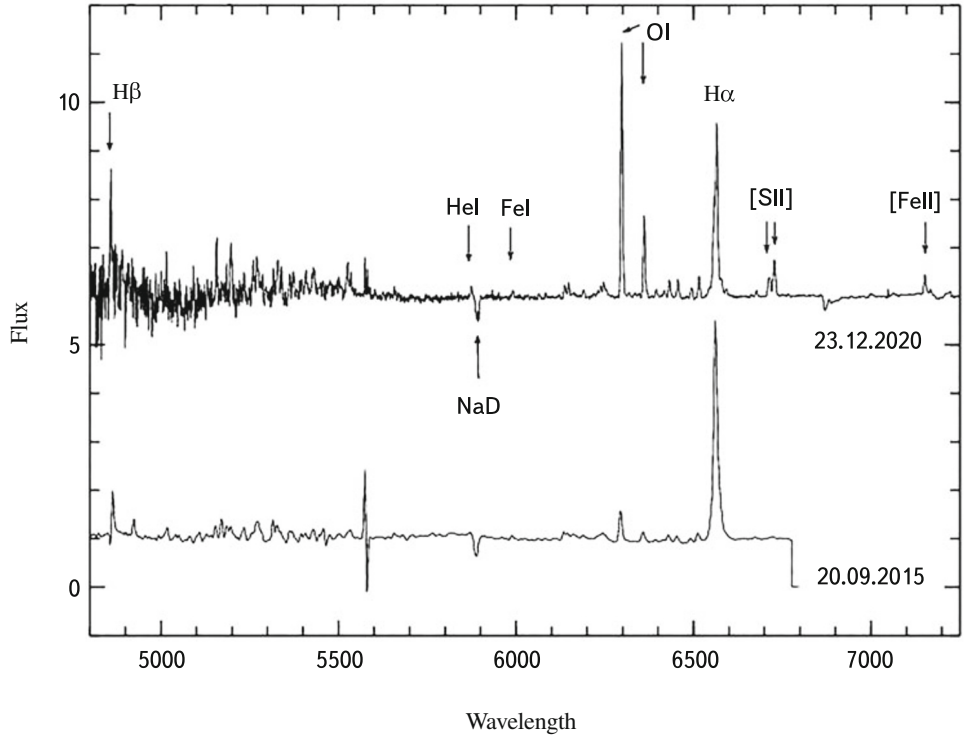

Fig. 2. Overall appearance of the spectrum of PV Cep on September 20, 2015 (at a maximum) and December 223, 2020 (at a minimum). For convenience the two traces are normalized to the continuum. The spectrum of December 23, 2020 is shifted upward by 5 nominal units.

Since the H and K lines of CaII did not fall in the spectral range of our spectra, we could not verify whether the fluorescence in the FeI lines is retained in the spectrum. Nevertheless, we note that the neutral iron lines $\lambda 6138$ FeI and $\lambda 6592.91$ FeI subject to fluorescence in the red region of the spectrum are constantly present, but their equivalent widths are lower by a factor of two than, for example, in the spectrum of the star Lk H$\alpha$ 120, with which PV Cep is often compared and where this effect is clearly expressed distinctly [24].

The most intense emission in the spectrum of PV Cep, the H$\alpha$ line, shows clearly noticeable profile changes that depend on the brightness level. As pointed out in essentially all of the previous studies, it consists of two components separated by a central absorption. We note that a two-peaked structure has also been observed in the first members of the Paschen series in the near IR [13]. Figure 3 shows sections of our spectra normalized to the continuum that include the Hα line. The figure shows that as the brightness decreases, the central absorption component becomes fainter, while on June 10 and, especially, December 21, 2016 (right at the time of maximum brightness) it fell below the level of the continuum (the EW was about 0.8 Å). However, a really brightly expressed P Cyg profile was observed only during the period of the absolute maximum of 1977-78 [4].

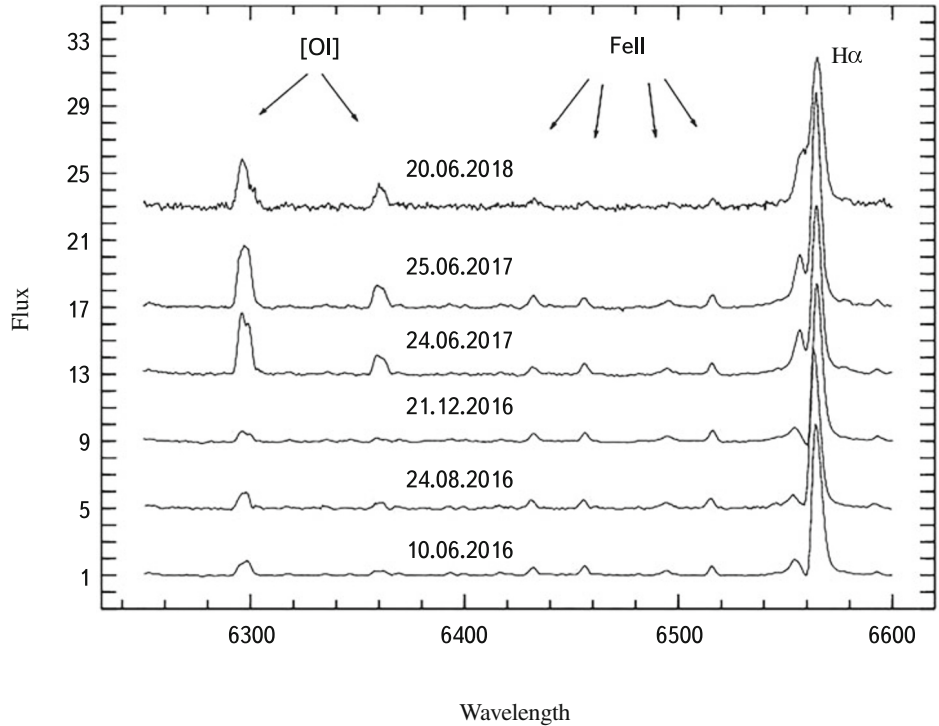

Fig. 3. Sections of the spectrum of PV Cep corresponding to different observation dates (d.m.y.) normalized to the continuum. Strong changes in the profiles and intensities of the Hα and [OI] lines can be seen, while the FeII emissions remain unchanged. The flux is plotted in arbitrary units on the ordinate; the spectra of 2016-2017 are shifted by 4 units relative to one another, and the spectrum of 2018 by 6 units.

A representation of changes of typical lines in the spectrum of PV Cep is given by the plots of the variation in the equivalent widths shown in Fig. 4. For a more intuitive comparison of these changes with the level of the visible brightness of the star, in Fig. 5 we show the dependence of the equivalent widths of the Hα and [OI] emissions on m(R). The errors in measuring the equivalent widths in our case depend mainly on the accuracy of taking the continuum spectrum and we estimate that they do not exceed 10-15%. It may be concluded from Figs. 4 and 5 that the equivalent widths of the forbidden lines unambiguously increase as the brightness decreases, while the equivalent width of one of the strongest permitted emissions (λ6517 Å Fe I) essentially does not change. The Hα line behaves in an ambiguous manner. The authors of earlier studies came to the same conclusions. It should be noted that, according to Ref. 12, the equivalent width of Hα in 2004-2009 was extremely high (>100 Å), essentially the same as during the brightness maximum of 1977-1979 (see the summary graph in Ref. 10; in subsequent years it does not reach these values).

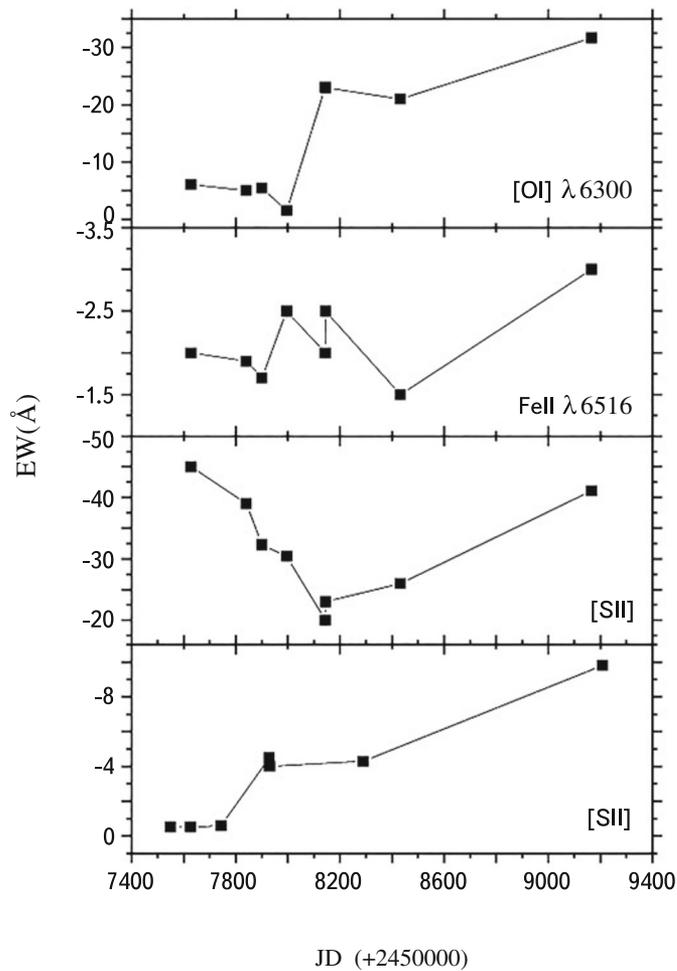

Fig. 4. Time variation in the equivalent widths of emission lines in the spectrum of PV Cep. For the Hα line the EW of the emission as whole was measured; for the [SII] lines the combined EW of the two doublet lines is shown.

**3.3. Radial velocities.** Already during the first observations it became obvious that PV Cep is surrounded by an expanding shell. In addition, it gradually became clear that the star is the source of a bipolar outflow. Thus, the question of whether lines in its spectrum belong to one or another emitting region (active spots on the surface, inner disk, outflowing matter, a collimated jet) is fairly important. Radial velocities can clarify this picture. To compare our data with previous measurements we reduced all the results to heliocentric radial velocities, which are used below.

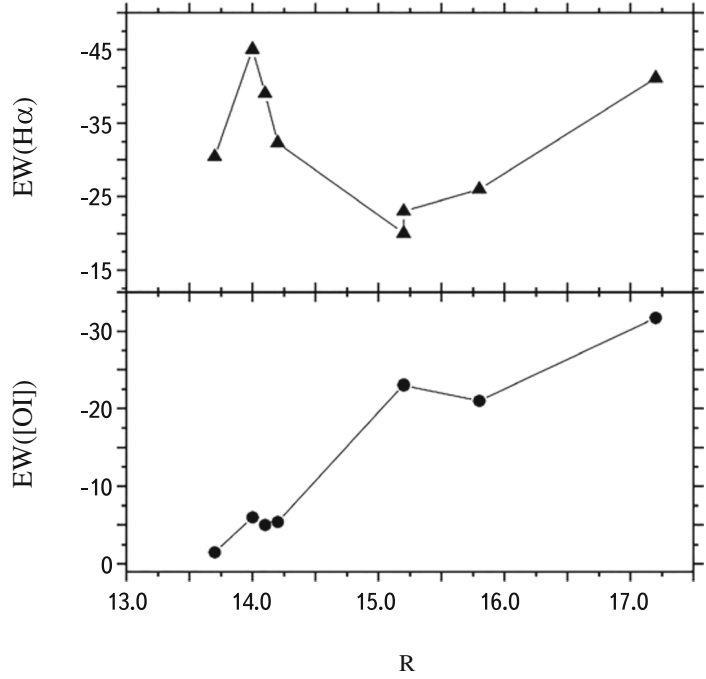

Fig. 5. The relationship between the visible brightness of PV Cep in the R band and the equivalent widths of the Hα and λ6300Å [OI] emission lines.

Since photospheric absorptions are not observed in the spectrum of PV Cep, the closest to the radial velocity of the star itself among the visible lines should be the FeII and FeI lines formed close to the surface. We obtained an average velocity of five FeII and FeI lines of -15±11 km/s, in nice agreement with the measurements of Herbig (-19 km/s) given in Ref. 4 and an estimate of -13 km/s in Ref. 25, as well as with the near zero values for the velocities of the permitted emissions in the near IR [25,13].

Emission components of the Hα line are evidently formed mainly in the expanding envelope created during the matter outflow. We note an extremely large total width of the emission (our estimates yield FWZI on the order of 1200-1500 km/s), as indicated by other authors. As can be seen from Fig. 3, ir does not depend on the brightness level, like the velocity of the main emission component (84±19 km/s). The absorption component of Hα, on the other hand, varies both in intensity (from a P Cyg profile to almost complete disappearance) and in radial velocity. Estimates based on the spectra of 2016-2018 give values from -120 to -180 km/s (note that during the epoch of the maximum of 1977 still larger shifts were observed [25]). For the minimum brightness of the star at the end of 2020, the shift in the Hα absorption was only -80 km/s.

The strong and narrow NaD absorptions, as for many similar stars, are also undoubtedly formed within an expanding envelope. It is remarkable that their intensity varies quite substantially, up to complete disappearance, but does not depend on the level of the visible brightness. Their radial velocity always remains negative, but its absolute magnitude varies, going beyond the limits of error of the measurements. In particular, in our spectra of 2016-2017 it ranged from -118 to -159 km/s and at the end of 2020 it equaled -45 km/s, i.e., it decreased like the Hα absorption.

These variations indicate that the intensity and outflow velocity of matter fluctuate quite substantially and this is probably related to the changes in the star's visible brightness.

**3.4. Forbidden lines.** Strong and broad [OI] and [SII] forbidden lines that are typical, as found later, for objects with a collimated outflow were detected in the spectrum of PV Cep immediately after its discovery. Their multicomponent structure was first described in Ref. 25. Not only the overall equivalent width of these lines change, but so do their radial velocities and the relative intensities of the components. These effects are especially distinct for the [SII] lines [10]. With sufficient resolution, in each of the lines of the doublet it is often possible to distinguish four components; on the other hand, sometimes they merge into one very wide (>400 km/s) emission band. The [OI] line also is divided into components, but because of the high intensity and width they are less expressed.

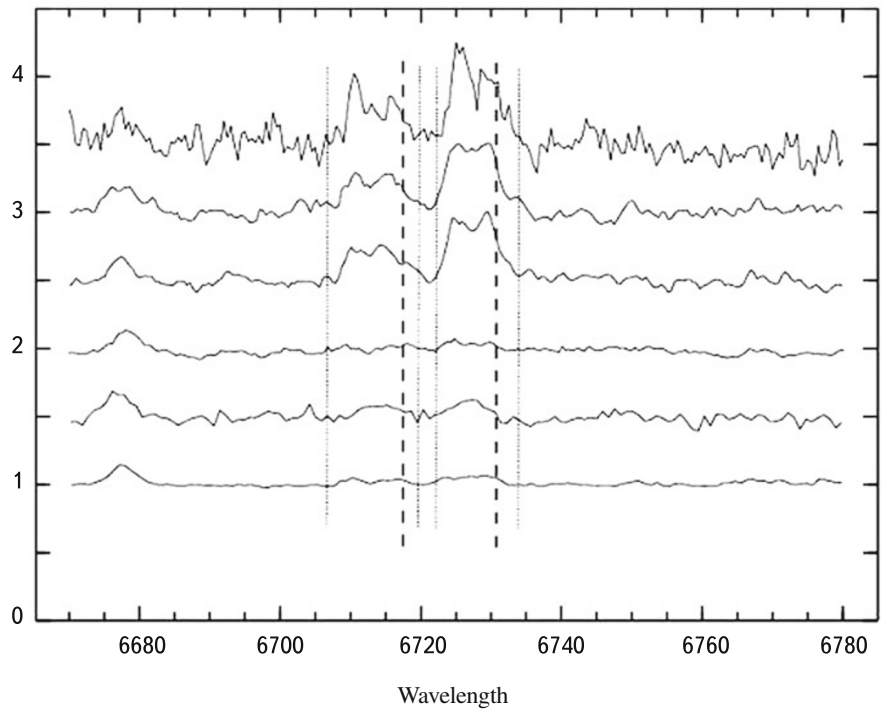

Fig. 6. Form and variation of the [SII] lines in the spectrum of PV Cep during 2016-2018. The values of the wavelengths corresponding to zero radial velocity are indicated by a dashed line. The limits for the FWZI of the lines are also indicated. The spectra are normalized to the continuum and shifted relative to one another along the ordinate by 0.5 arbitrary units of the flux.

These features make it more difficult to obtain accurate measurements, but on the average the radial velocities of the components of [SII] in the spectra examined in this paper the following values can be estimated as: -230, -170, -50, and +70 km/s (here the component with a positive velocity changes especially strongly and often is entirely absent). Their form is shown schematically in Fig. 6.

Taking into account aforesaid, on comparing these results with data from earlier work, we find very good agreement. Thus, in Ref. 25 components with -246, -133, and -55 km/s are identified in the [OI] line; in Ref. 26, values of -275 and -60 km/s are given (for an overall line width from -455 to +255 km/s) and for [SII], -300 and -75 km/s; and, in Ref. 27 for [OI[ for an overall range from -400 to +50 km/s components with -300, -170, -100, and -50 km/s are listed. Thus, components with a positive radial velocity are rarely observed, evidently because of their faint intensity.

The detection of similar split among the strong emission lines of [FeII], which are fairly numerous in the near IR (with strong changes in their intensity), is extremely important for understanding this effect [13]. These components show up quite clearly for the 1.257 and 1.644 mm lines of [FeII], with radial velocities of -265, -99, +50, and +165 km/s. In that same paper it was shown that the emission from components with a high absolute velocity actually belong to a collimated flow emerging from PV Cep while those with a low one are concentrated around the star.

Thus, it is logical to assume that the multicomponent structure of the [OI] and [SII] lines is of the same nature. However, as we see, a really good agreement in the velocities between them and [FeII] is observed only for the high-velocity blue-shifted component, which is also the least of all subject to intensity fluctuations; on the other hand, a high-velocity red-shifted component has never been observed in the visible.

This circumstance is quite naturally explained by a combination of two factors. First, in the immediate neighborhoods of PV Cep there are constant and rather brief variations in the circumstellar absorption caused by the formation and redistribution of dust particles in the disk during the outflow process and individual ejections of matter [12]. The average value of this absorption estimated from the [FeII] lines is of order $A_v$ = 10 [13]. The significant influence of circumstellar dust clouds is also indicated by the dependence of the equivalent width of the H$\alpha$ emission on the brightness of PV Cep (Fig. 5), which has features typical for UX Ori stars [28]. Since the inclination of the axis of the circumstellar disk of PV Cep to the sky plane is insignificant [29], it is logical to assume that near the star the emission of the red-shifted (i.e., away from us) branch of the flow in the visible is just screened by this disk and becomes well resolvable only in the near IR. Second, as it is reasonably suggested in Ref. 13, the small (2"-3") emission knot (actually a miniature jet) seen by the authors might be ejected from PV Cep during the outburst of 2004-2005 recorded in Ref. 12. This makes one to suspect that even for lower amplitude manifestations of activity around the star (at least along the axis of an outflow), small condensations appear emitting in forbidden lines. The differences in their densities, velocities, and sizes also create fairly rapidly varying profiles of the [OI] and [SII] lines.

A logical conclusion from the above discussion is the assumption that even after the brightness maximum between 2016 and 2017 described in section 3.1, one could expect the appearance of a similar knot that might be observed directly in coming years (and thereby confirming this hypothesis). Analogous observations of the formation of new clumps in jets after outbursts are extremely rare and have been noticed, as far as we know, only for a companion of Z CMa [30] and a newly discovered fuor V2494 Cyg [31]. As an object undergoing fairly frequent and powerful outbursts, in this sense PV Cep opens up good possibilities for testing and careful study of this mechanism.

**4. V350 Cep**

The stars PV Cep and V350 Cep were not only discovered simultaneously, but are rather similar in terms of the overall form of their spectra, so it is entirely logical to compare these two objects.

As already noted above, since 1978 V350 Cep has essentially been in a maximum the entire time; accordingly, the spectrum that we obtained and have examined in this paper belongs to a quiescent period in the state of the star. In general, it is completely consistent with the results of previous studies. Some fragments are shown in Fig. 7.

The measured equivalent widths of the most characteristic lines of FeI, FeII, and HeI were in excellent agreement with the values of 1990 given in Ref. 19. The intensity of the FeI fluorescence lines in our spectrum is slightly lower than during the 1978-1994 period [20]. The intensities of the forbidden [OI] and [SII] lines also were essentially unchanged. The Hα line in our spectrum of 2015 has a distinct P Cyg profile (Fig. 7). In terms of equivalent width (about -24 Å for the main emission component and 0.2 Å for the absorption component, which falls below the continuum) it corresponds to the data of 1990, 1994, 2005, and 2007 [19,6,21]. The radial velocities of the forbidden emissions turned out to be negative: -117 km/s for [OI] and -97 km/s for [SII]. For the Hα line, values of +46 and -270 km/s (emission peaks) and -167 km/s (absorption component) were obtained. These values for Hα

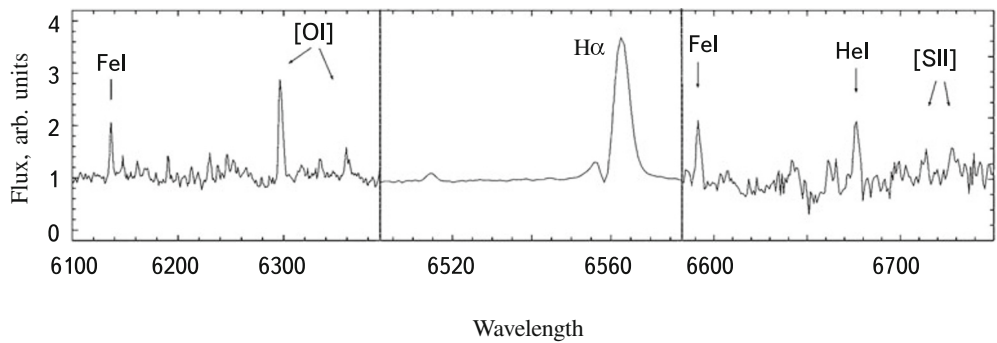

Fig. 7. Fragments of the spectrum of V350 Cep normalized to the continuum which manifest a P Cyg profile of the Hα line, forbidden [OI] and [SII] lines, and fluorescence lines of FeI.

in very good agreement with the results of 2005 and 2007.

Thus, the activity of V350 Cep has been essentially at an unchanged level since the time of the rise in brightness, i.e., for more than 40 years. A comparison of available published data on the profile of the Hα line, however, leads us to a curious conclusion that has not been mentioned anywhere before. Namely, from the time of the first observations (1978) through July 1985 [18,32,20] no P Cyg profile was observed in the Hα line, while since October 1986 it has been constantly present [20,19,6,21, and this paper].

The above-mentioned facts indicate that V350 Cep is surrounded by an accretion disk which creates a far IR excess and sometimes shows up in UX Ori-like brightness drops [33]. However, the presence of a disk does not strongly affect the extinction: $A_v$ = $1^m.2$ [34]. After the brightness of a star increased, a fairly dense expanding envelope was formed with a range of velocities of several hundred km/s and because of a gradual increase in densities in it, P Cyg-type profiles showed up. The question of the existence of a collimated flow in V350 Cep remains open. On one hand, such a flow was not been detected by direct observations (the results of Ref. 19 are not entirely convincing); on the other hand, it is possible that V350 Cep may be the source of the Herbig-Haro object HH 235 [35], although the rather faint forbidden [OI] and [SII] lines indicate, in any case, that the intensity of shock excitation in the shell of the star itself is insignificant.

5. Discussion

The absence of any kind of photospheric lines in the spectrum of PV Cep makes it extremely difficult to classify it and determine other parameters. A spectral type A was estimated in Ref. 4 rather arbitrarily. The only case where absorption lines corresponding to a later type (G8-K0) were observed is mentioned in Ref. 10 when in 1978 absorption lines were observed in the blue region of the PV Ce spectrum, but since a spectrum of this kind was never observed again, this estimate should also be dealt with cautiously.

The estimated distances of PV Cep have also differed (by authors mostly from different stellar astronomy studies): 500 pc [4], 440 pc [27], 280 pc [34], and 325 pc [12]. At present, due to the availability of data from the Gaia observatory, the parallax of PV Cep is known very precisely; it corresponds to a distance of 340 pc [36].

Estimating the absorption in the direction to PV Cep is also a fairly difficult problem. The color of a star and a rapid drop in the intensity of the continuum in the blue region indicate only that the absorption is substantial. The most of it undoubtedly arises in the circumstellar disk. It changes noticeably with time, with a significant influence on the observed fluctuations in the brightness of PV Cep. In addition, as shown in Ref. 13, a substantial difference is observed in the absorption between the northern and southern (the emission of which passes through the disk) fractions of the emerging outflow.

It is not surprising that estimating the bolometric luminosity of PV Cep is a complicated problem. Considering the latest work, in Ref. 12 the bolometric luminosity of the star itself was estimated to be $\approx 17 L_\odot$ (for a spectral class of G8 taken from Ref. 10), while the luminosity of the accretion disk, which also creates a substantial fraction of the total emission of PV Cep, turned out to be substantially higher: $\approx 40\,L_\odot$ for the low brightness level and $\approx 80\,L_\odot$ at the maximum of 2006. However, in Ref. 13 the estimate of the accretion luminosity based on the data of 2012 was an order of magnitude lower at $\approx 5.7\,L_\odot$, which the authors explain by a further reduction in the accretion rate. If we note that a clearly excessive distance to the object (500 pc for an actual 340 pc) is used in Ref. 13, this discrepancy seems even greater. Thus, we may assume that in the current state of a deep minimum, the brightness of PV Cep is basically determined just by the emission from the star itself, i.e., $L_{\text{bol}} \approx 20\,L_\odot$.

The situation with V350 Cep, the spectral class of which is reliably known, is simpler. The distance to the cluster NGC 7129 which includes this star has been estimated by various authors to be from 800 to 1260 pc. However, the parallaxes obtained with the Gaia space observatory give a fairly accurate value of 900 pc [37]. Estimating an average value $16^m.1$ [16] of V for V350 Cep from the light curve for 1993-2014 and taking the above values of distance and absorption, as well as the bolometric correction for spectral class M2 [38], we obtained an estimate of $3.3\,L_\odot$ for the post-outburst luminosity of this object. Unfortunately it was not possible to separate the accretion luminosity component using the available data.

The main question, i.e., the unusual character of the variability of the two stars and their classification, remains open as before. We first examine V350 Cep. It is evident that the rise in the star's brightness is caused by a sharp increase in the accretion rate, but the subsequent stable state at the maximum level is unique of its kind. Yet another recently published historical light curve for 1970-1985 confirms that the rise in the brightness began right in 1971-72 and no other significant outbursts have occurred since 1985 [33].

V350 Cep was initially assigned to the class of possible EXors [5], then excluded from it [6], first of all because of the absence of recurrent outbursts (although it corresponds to typical EXors in terms of luminosity). On the other hand, according to the light curve this object corresponds exactly to classical FUors with slowly rising luminosity, e.g., V1515 Cyg, and after 1997 a very slow reduction in brightness characteristic for many FUors has been observed [16]. A possible relationship to a collimated outflow also relates V350 Cep to the FUors. The impressive stability of the spectral characteristics over almost 40 years (except for the Hα line) is also noteworthy. P Cyg components in the Hα and NaD lines with a broad and almost rectangular profile also resemble the FUors [6,21]. The rather low luminosity of V350 Cep is not typical for the FUors, but is still not exclusive. Otherwise, however, the abundant emission spectrum of V350 Cep corresponds to CTTS and not to FUors. We tend to agree with the assumption [16] of a similarity between V350 Cep and V1647 Ori, a recently discovered eruptive object that combines features of both a FUor and an EXor.

The observational facts collected up to now make it possible with sufficient confidence to conclude that there is an entire class of intermediate objects between FUors and EXors, to which almost twenty objects may be assigned (tentatively), of which half are only visible in the IR. Overall, V 350 Cep also is one of them. If subsequent observations confirm the stability or further intensification of the P Cyg component in the Hα line, it may be possible to assume that V350 Cep is gradually continuing to develop toward a typical FUor state. In conclusion, we note that important information for classification of FUors and FUor-like objects can be obtained from infrared spectra. As far as we know, IR spectroscopy of V350 Cep has not yet been carried out.

PV Cep, which unlike V350 Cep, manifests repeated eruptions, also cannot be assigned to EXors for a whole series of reasons, that are examined in detail in Refs. 39 and 13. Here we note only the basic reasons: PV Cep is an object with a higher mass and luminosity than classical EXors. It is surrounded by a fairly massive accretion disk. Calculations for the time of and immediately after the outburst [40, 12] give an accretion rate of roughly $10^{-6}$ $M_\odot$ yr$^{-1}$, and it fell subsequently to $10^{-7}$ $M_\odot$ yr$^{-1}$ [13], but even this value is higher than in classical CTTS stars. PV Cep is undoubtedly in an earlier stage of evolution than the EXors. Finally, it is associated with a biconical cometary nebula and generates a powerful, extended (2.6 pc) collimated outflow [7,8] which is utterly atypical for EXors.

It is extremely noteworthy, however, that almost all the properties enumerated above are, on the other hand, typical of FUors. PV Cep, in turn, differs from the classical fuors in having a highly developed emission spectrum and, of course, repeated outbursts. Thus, like V350 Cep it is an object that resembles V 1647 Ori, which is intermediate between the FUors and the EXors. The number of such "intermediate" objects has increased in recent years; in particular, V346 Nor, V2492 Cyg, and V1180 Cas may be among them. We plan to return to the questions of classifying young eruptive stars in the future.

We are grateful to the observers at AAVSO for providing data of invaluable significance for the study of variable stars. The observations at the telescope of the Special Astrophysical Observatory were supported by the Ministry of Science and Higher Education of the Russian Federation (including agreement No05.619.21.0016, unique project identifier RFMEFI61919X0016).

We thank a reviewer for valuable advice and comments.

This work was partially supported by the Committee on Science of the Republic of Armenia as part of scientific project 18T-1C329.